\documentclass[fleqn,11pt,twoside]{article}

\usepackage{amsthm,amsthm,amssymb, color, xcolor,epsfig, graphics, subfigure}

\usepackage{amsmath, graphicx, latexsym, lscape }

\makeatletter
\newcommand{\copyrightnote}[2]{{\renewcommand{\thefootnote}{}
 \footnotetext{\small\it
\begin{flushleft}
 \copyright \ #1   #2
\end{flushleft}}}}

\newcommand{\Name}[1]{\begin{flushleft}
                       \LARGE \bf #1
                       \end{flushleft}\vspace{-3mm}}

\newcommand{\Author}[1]{\begin{flushleft}
                       \it #1 \end{flushleft}}

\newcommand{\Address}[1]{\begin{flushleft}
                       \it #1 \end{flushleft}}

\newcommand{\Date}[1]{\begin{flushleft}
                      \small  \it #1 \end{flushleft}}

\newcommand{\evenhead}{Author \ name}
\newcommand{\oddhead}{Article \ name}

\renewcommand{\@evenhead}{
\hspace*{-3pt}\raisebox{-15pt}[\headheight][0pt]{\vbox{\hbox to \textwidth
{\thepage \hfil \evenhead}\vskip4pt \hrule}}}
\renewcommand{\@oddhead}{
\hspace*{-3pt}\raisebox{-15pt}[\headheight][0pt]{\vbox{\hbox to \textwidth
{\oddhead \hfil \thepage}\vskip4pt\hrule}}}
\renewcommand{\@evenfoot}{}
\renewcommand{\@oddfoot}{}

\long\def\@makecaption#1#2{%
  \vskip\abovecaptionskip
  \sbox\@tempboxa{\small \textbf{#1.}\ \ #2}%
  \ifdim \wd\@tempboxa >\hsize
    {\small \textbf{#1.}\ \ #2}\par
  \else
    \global \@minipagefalse
    \hb@xt@\hsize{\hfil\box\@tempboxa\hfil}%
  \fi
  \vskip\belowcaptionskip}

\newcommand{\JNMPnumberwithin}[3][\arabic]{%
  \@ifundefined{c@#2}{\@nocounterr{#2}}{%
    \@ifundefined{c@#3}{\@nocnterr{#3}}{%
      \@addtoreset{#2}{#3}%
      \@xp\xdef\csname the#2\endcsname{%
        \@xp\@nx\csname the#3\endcsname .\@nx#1{#2}}}}%
}

\newcommand{\resetfootnoterule} {
  \renewcommand\footnoterule{%
  \kern-3\p@
  \hrule\@width.4\columnwidth
  \kern2.6\p@}
}

\renewcommand{\footnoterule}{}

\makeatother

\theoremstyle{definition}

\newtheorem{theorem}{Theorem}[section]

\begin{document}

\renewcommand{\evenhead}{ {\LARGE\textcolor{blue!10!black!40!green}{{\sf \ \ \ ]ocnmp[}}}\strut\hfill S.V. Meleshko and C. Rogers}
\renewcommand{\oddhead}{ {\LARGE\textcolor{blue!10!black!40!green}{{\sf ]ocnmp[}}}\ \ \ \ \   Reciprocal Transformations and
Lie Group Connections}

\thispagestyle{empty}
\newcommand{\FistPageHead}[3]{
\begin{flushleft}
\raisebox{8mm}[0pt][0pt]
{\footnotesize \sf
\parbox{150mm}{{Open Communications in Nonlinear Mathematical Physics}\ \  \ \ {\LARGE\textcolor{blue!10!black!40!green}{]ocnmp[}}
\quad Vol.1 (2021) pp
#2\hfill {\sc #3}}}\vspace{-13mm}
\end{flushleft}}

\FistPageHead{1}{\pageref{firstpage}--\pageref{lastpage}}{ \ \ Article}

\strut\hfill

\strut\hfill

\copyrightnote{The author(s). Distributed under a Creative Commons Attribution 4.0 International License}

\Name{Reciprocal Transformations in Relativistic Gasdynamics.
Lie Group Connections}

\Author{Sergey V. Meleshko$^{\,1}$ and Colin Rogers$^{\,2}$}

\Address{$^{1}$ School of Mathematics, Institute of Science, Suranaree University of Technology, Thailand\\[2mm]
$^{2}$ School of Mathematics and Statistics,  The University of New South Wales, Sydney, NSW2052,
Australia}

\Date{Received April 14, 2021; Accepted July 13, 2021}

\setcounter{equation}{0}

\begin{abstract}
\noindent
Reciprocal transformations associated with admitted conservation laws were originally
used to derive invariance properties in non-relativistic gasdynamics and
applied to obtain reduction to tractable canonical forms. They have subsequently been
shown to have diverse physical applications to nonlinear systems, notably in the analytic treatment of Stefan-type moving boundary problem and in linking inverse scattering systems and integrable hierarchies in soliton theory. Here,invariance under classes of reciprocal
transformations in relativistic gasdynamics is shown to be linked to a Lie group procedure.

\end{abstract}

\label{firstpage}


\section{Introduction}

Bateman \cite{hb38} in a study of lift and drag aspects in two-dimensional homentropic irrotational gasdynamics established invariance of the governing system under a novel multi-parameter class of relations which have come to be known as reciprocal transformations. The latter, characteristically are associated with conservation laws admitted by a system. The reciprocal transformations of \cite{hb38} were subsequently discussed by Tsien \cite{ht39} in connection with the application of the K\'arm\'an-Tsien model pressure-density law which, in turn, has its roots in the classical work of Chaplygin \cite{sc33} on gas jets. Thus, with a K\'arm\'an-Tsien law, the class of reciprocal transformations derived by Bateman may be applied to link, in subsonic r\'egimes, a gasdynamic system with an associated hydrodynamic system encapsulated in a Cauchy-Riemann system. Loewner \cite{cl52} later in this gasdynamics context sought via infinitesimal B\"acklund transformations to construct systematically model gas laws which asymptotically lead to such a canonical reduction. It is remarkable that contained therein, \textit{mutatis mutandis} is a linear representation to be re-discovered some twenty years later for the sine Gordon equation. It accordingly constitutes a seminal connection between modern soliton theory and the subsequent applications therein of B\"acklund transformations \cite{crws82,crws02}.

It was established in \cite{crwskc07} that the kind of infinitesimal transformations introduced by Loewner in \cite{cl52} may likewise be applied in Lagrangian nonlinear elastodynamics. In that case, novel model stress-strain laws were derived which are linked to an integrable sinh-Gordon equation and which, importantly, admit an interior change of concavity. This material phenomenon can occur, for instance, in the compression of polycrystalline materials \cite{jkev74} and in the unloading r\'egimes of superelastic nickel-titanium \cite{tdapds96,td95}.

In \cite{bkcr91,bkcr93}, a re-interpretation and extension of the infinitesimal B\"acklund transformations of \cite{cl52} was shown to lead to a linear representation for a wide class of 2+1-dimensional solitonic equations. The $\bar{\partial}$-dressing method was outlined for these generalised Loewner systems. In particular, a novel canonical 2+1-dimensional sine Gordon equation was derived which, like the Davey-Stewartson and Nizhnik-Novikov-Veselov equations contain the two spatial variables with equal standing.

In soliton theory, reciprocal transformations associated with admitted conservation laws and their incorporation in Bianchi diagrams linked to auto-B\"acklund transformations were introduced in \cite{jkcr82}. Reciprocal transformations in 1+1-dimensions have been applied in the linkage of the AKNS and WKI inverse scattering schemes \cite{crpw84} and in a range of solitonic contexts (see e.g. \cite{crsc87}$-$\cite{ktql12} and literature cited therein). In \cite{wocr93} reciprocal transformations in 2+1-dimensions were applied to connect the Kadomtsev-Petviashvili, modified Kadomtsev-Petviashvili and Dym integrable hierarchies. The links therein were formulated as Darboux theorems for associated Lax operators.

In nonlinear continuum mechanics, reciprocal transformations have likewise proved to have diverse physical applications. In \cite{cr85,cr86}, they have been applied to solve moving boundary problems which arise in the modelling of melting in a range of metals as described in the seminal work of Storm \cite{ms51}. Nonlinear moving boundary problems arising in sedimentation theory \cite{crpb92}, the percolation of liquids through porous media such as soils \cite{crpb88}, and methacrylate saturation processes \cite{afcrws05} have likewise been solved via the application of reciprocal transformations. In recent work, reciprocal transformations have been applied in \cite{cr15} in connection with moving boundary problems associated with the solitonic Dym equation. It is recalled that a link has been established in \cite{lk90,gvlk91} between a classical Saffman-Taylor problem with interface moving at constant speed and certain solutions of the canonical Dym equation \cite{pv01}.

The preceding attests to the importance of reciprocal transformations in physical applications. In \cite{art:IbragimovRogers2012}, a novel link was established between a one-parameter subclass of infinitesimal reciprocal-type transformations in gasdynamics and a Lie group approach. Thereby, a standard one-parameter $(\beta)$ class of reciprocal transformations may be reconstructed via a standard Lie group procedure involving solution of a system of initial value problems. The identity transformation is retrieved in the limit $\beta\rightarrow\infty$. In \cite{cryh12}, this one-parameter class of reciprocal transformations was applied to a 1+1-dimensional Lagrangian anisentropic system with a Prim-type gas law associated with the affinsph\"aren equation which arises in the theory of Tzitzeica surfaces \cite{wscr94,crwskccm05}. An integrable deformed version of this solitonic affinsph\"aren equation was obtained thereby and a linear representation set down.

In \cite{art:RogersRuggeri2020,art:RogersRuggeriSchief2020}, invariance under multi-parameter reciprocal transformations has recently been established in relativistic gasdynamics. Here, a connection between one-parameter subclasses of such transformations and a Lie group procedure is established.

\section{Reciprocal invariance in 1+1-dimensional relativistic gasdynamics}

In \cite{cr68}, a reciprocal invariance property was established as embodied in the following:

\begin{theorem} The relativistic 1+1-dimensional gasdynamic system
\begin{align}
\begin{aligned}\frac{\partial}{\partial t}\left(\frac{\rho c}{\sqrt{c^{2}-v^{2}}}\right)+\frac{\partial}{\partial x}\left(\frac{\rho cv}{\sqrt{c^{2}-v^{2}}}\right)=0,\\
\frac{\partial}{\partial t}\left(\frac{(e+p)v}{c^{2}-v^{2}}\right)+\frac{\partial}{\partial x}\left(\frac{(e+p)v^{2}}{c^{2}-v^{2}}+p\right)=0,
\end{aligned}
\label{eq:Mar24.1}
\end{align}
is invariant under the 4-parameter class of reciprocal transformations
\begin{equation} \label{b2}
\left.\begin{array}{c}
\rho^{*}=\dfrac{a_{3}\rho\sqrt{(p+a_{2})^{2}-a_{1}^{2}v^{2}/c^{2}}}{\sqrt{1-v^{2}/c^{2}}}\left[\dfrac{1}{p+((e+p)/c^{2})v^{2}/(1-v^{2}/c^{2})+a_{2}}\right]\ ,\\[6mm]
v^{*}=-\dfrac{a_{1}v}{p+a_{2}}\ ,\\[6mm]
e^{*}=\dfrac{a_{3}S(p+a_{2})(1-a_{1}^{2}v^{2}/c^{2}(p+a_{2})^{2})}{p+Sv^{2}+a_{2}}c^{2}-a_{4}+\dfrac{a_{1}^{2}a_{3}}{p+a_{2}}\ ,\\[6mm]
p^{*}=a_{4}-\dfrac{a_{1}^{2}a_{3}}{p+a_{2}}
\end{array}\right\} \ \mathbb{R}^{*}
\end{equation}
with
\begin{equation} \label{b3}
a_{1}dt^{*}=Svdx-(p+Sv^{2}+a_{2})dt\ ,\quad dx^{*}=dx
\end{equation}
and $S=(e+p)/(c^{2}-v^{2}). \qquad \square$

\end{theorem}

The above result in the limit $c\rightarrow\infty$ reproduces the 4-parameter class of reciprocal invariant transformations originally obtained in 1+1-dimensional non-relativistic gasdynamics in \cite{cr68}. Cognate adjoint-type invariant transformations have application in the analysis of gas flow between a piston and a non-uniform shock in \cite{sccr74}.

In \cite{art:RogersRuggeri2020}, a one-parameter sub-class of reciprocal transformations which constitute a specialisation of those in \cite{cr68} was reconstructed via application of Lie group methods involving an associated Cauchy initial value problem. Here, the subclass of reciprocal transformations \eqref{b2}--\eqref{b3} for the system \eqref{eq:Mar24.1} is considered with
\begin{equation} \label{b4}
\begin{array}{c} a_1=-\epsilon^{-1}\ , \quad a_2=a_4=\epsilon^{-1}\\[1mm]
a_3=1 \end{array} \end{equation}
and accordingly becomes
\begin{equation}
\begin{array}{c}
\rho^{*}=\dfrac{\rho\sqrt{(\epsilon p+1)^{2}-v^{2}/c^{2}}}{\left(\epsilon\left(p+((e+p)/c^{2})v^{2}/(1-v^{2}/c^{2})\right)+1\right)\sqrt{1-v^{2}/c^{2}}}\ ,\\[3pt]
v^{*}=\dfrac{v}{\epsilon p+1}\ ,\,\,\,p^{*}=\dfrac{p}{\epsilon p+1}\\[3pt]
e^{*}=\dfrac{S(c^{2}(\epsilon p+1)^{2}-v^{2})}{(\epsilon(p+Sv^{2})+1)(\epsilon p+1)}c^{2}-\dfrac{p}{\epsilon p+1}\ ,\\[6mm]
\end{array}\label{eq:Mar29.1}
\end{equation}

\begin{equation}
dt^{*}=-\epsilon\left(Svdx-(p+Sv^{2})dt\right)+dt\ ,\quad dx^{*}=dx\label{eq:mar27.1}
\end{equation}

It is readily verified that the one-parameter class of relations \eqref{eq:Mar29.1}-\eqref{eq:mar27.1} constitute a Lie group.
Moreover, the reciprocal relations \eqref{eq:Mar29.1}-\eqref{eq:mar27.1} satisfy the Cauchy problem
\begin{equation}
\begin{array}{rcll}
{\displaystyle \frac{d\rho^{*}}{d\epsilon}} & = & -\rho^{*}v^{*2}e^{*}\Gamma^{*2}, & \,\,\,\rho_{|\epsilon=0}^{*}=\rho,\\[2ex]
{\displaystyle \frac{dv^{*}}{d\epsilon}} & = & -p^{*}v^{*}, & \,\,\,v_{|\epsilon=0}^{*}=v,\\[2ex]
{\displaystyle \frac{dp^{*}}{d\epsilon}} & = & -p^{*2}, & \,\,\,p_{|\epsilon=0}^{*}=p,\\[2ex]
{\displaystyle \frac{de^{*}}{d\epsilon}} & = & (c^{2}p^{*2}-v^{*2}e^{*2})\Gamma^{*2}, & \,\,\,e_{|\epsilon=0}^{*}=e,\\[2ex]
{\displaystyle \frac{d(dt^{*})}{d\epsilon}} & = & S^{*}v^{*}dx^{*}-(p^{*}+S^{*}v^{*2})dt^{*}, & \,\,\,(dt^{*})_{|\epsilon=0}=dt,\\[2ex]
{\displaystyle \frac{d(dx^{*})}{d\epsilon}} & = & 0, & \,\,\,(dx^{*})_{|\epsilon=0}=dx\ ,
\end{array}\label{eq:Mar29.1-1}
\end{equation}
where $\Gamma^*=(c^2-v^{*2})^{1/2}$.
Similar to the classical theory of a Lie group of point transformations,
it is convenient to present them by the
infinitesimal generator
\begin{equation}
X=-\rho v^{2}e\Gamma^{2}\partial_{\rho}-pv\partial_{v}-p^{2}\partial_{p}+(c^{2}p^{2}
-e^{2}v^{2})\Gamma^{2}\partial_{e}+\left((c^{2}p+ev^{2})dt
-v(e+p)dx\right)\Gamma^{2}\partial_{dt}.
\label{eq:generator_original}
\end{equation}

An advantageous aspect of this infinitesimal approach
is that it reveals hidden invariants under the reciprocal
transformations. In this connection, a function $J$ is an invariant of the group of transformations
(\ref{eq:Mar29.1}), (\ref{eq:mar27.1}) if and only if it satisfies
the differential equation \cite{bk:Ovsiannikov1978}
\[
XJ=0.
\]
A particular triad of such invariants is given by
\begin{equation} \label{b9}
\begin{array}{c}
J_{1}=\dfrac{(cp-ve)(c-v)}{(cp+ve)(c+v)},\,\,\,J_{2}=\dfrac{\rho p}{(cp+ev)}\sqrt{\dfrac{c-v}{c+v}},\\[5mm]
J_{3}=\dfrac{v(cp+ev)}{p(c-v)(pc^{2}+ev^{2})}\left((c^{2}p+ev^{2})dt-v(e+p)dx\right).
\end{array} \end{equation}

It is remarked that the above procedure to construct one-parameter invariant transformations may, in principle, be extended by application of the infinitesimal generator
\begin{equation} \label{b10}
Y=\zeta^{\rho}\partial_{\rho}+\zeta^{v}\partial_{v}+\zeta^{p}\partial_{p}+\zeta^{e}\partial_{e}+\left((c^{2}p+ev^{2})dt-v(e+p)dx\right)\Gamma^{2}\partial_{dt},
\end{equation}
wherein the functions $\zeta^{\rho}$, $\zeta^{v}$, $\zeta^{p}$ and
$\zeta^{e}$ depend on $(\rho,v,p,e)$. This is under present investigation.

\section{Reciprocal invariance in two-dimensional relativistic gas dynamics}

In \cite{art:RogersRuggeriSchief2020} the following invariance result was established:

\begin{theorem} \label{th_2}The two-dimensional relativistic gasdynamic
system
\begin{equation}
\begin{aligned}\partial_{x}(Ru)+\partial_{y}(Rv) & =0,\\
\partial_{x}(p+Su^{2})+\partial_{y}(Suv) & =0,\\
\partial_{x}(Suv)+\partial_{y}(p+Sv^{2}) & =0,\\
\partial_{x}(Su)+\partial_{y}(Sv) & =0,
\end{aligned}
\label{eq:Mar31.10}
\end{equation}
where
\begin{equation} \label{c2}
q^{2}=u^{2}+v^{2},\,\,\,\Gamma=\frac{1}{\sqrt{c^{2}-q^{2}}},\quad R=\rho\Gamma,\quad S=\frac{e+p}{c^{2}}\Gamma^{2},
\end{equation}
is invariant under the 4-parameter class of transformations
\begin{equation}
\begin{array}{c}
u^{*}=-\dfrac{a_{1}u}{p+a_{2}}\ ,\quad v^{*}=-\dfrac{a_{1}v}{p+a_{2}},\quad p^{*}=a_{4}-\dfrac{a_{1}^{2}a_{3}}{p+a_{2}},\\[5mm]
e^{*}=\dfrac{a_{3}c^{2}(p+a_{2})(e+p)\Delta}{(p+a_{2})(c^{2}-q^{2})+(e+p)q^{2}}-a_{4}+\dfrac{a_{1}^{2}a_{3}}{p+a_{2}}\\
{\displaystyle \rho^{*}=a_{3}\rho c\Gamma\sqrt{\Delta}\frac{p+a_{2}}{p+Sq^{2}+a_{2}},}
\end{array}\label{b26}
\end{equation}
where
\begin{equation}
\Delta=1-a_{1}^{2}\frac{q^{2}}{c^{2}(p+a_{2})^{2}},\quad S=(e+p)\Gamma^{2},\label{b27a}
\end{equation}
and
\begin{equation}
\begin{array}{c}
dx^{*}=-(p+Sv^{2}+a_{2})dx+Suv\,dy,\quad dy^{*}=Suv\,dx-(p+Su^{2}+a_{2})dy,\\[3mm]
0<|J(x^{*},y^{*}\ ;\ x,y)|<\infty.
\end{array}\label{b28}
\end{equation}
\end{theorem}

In the following, it proves convenient to introduce a scaling $x^*\rightarrow a_1x^*,\ y^*\rightarrow a_1y^*$ which leaves the preceding theorem unchanged.

Here, we proceed with the specialisation of parameters \eqref{b4} whence the corresponding subclass of
reciprocal transformations becomes
\begin{equation}
\begin{array}{c}
u^{*}=\dfrac{u}{\epsilon p+1}\ ,\quad v^{*}=\dfrac{v}{\epsilon p+1},\quad p^{*}=\dfrac{p}{\epsilon p+1},\,\,\,{\displaystyle \rho^{*}=\rho c\Gamma\sqrt{\tilde{\Delta}}\frac{\epsilon p+1}{\epsilon(p+Sq^{2})+1},}\\[5mm]
e^{*}=\dfrac{c^{2}(\epsilon p+1)(e+p)\tilde{\Delta}}{(\epsilon p+1)(c^{2}-q^{2})+\epsilon(e+p)q^{2}}-\dfrac{p}{\epsilon p+1},
\end{array}\label{eq:Mar31.15}
\end{equation}
where ${\displaystyle \tilde{\Delta}=1-\frac{q^{2}}{c^{2}(\epsilon p+1)^{2}}}$,
and

\begin{equation}
dx^{*}=\epsilon((p+Sv^{2})dx-Suv\,dy)dx+dx,\quad dy^{*}=\epsilon(-Suv\,dx+(p+Su^{2})dy)+dy,\label{eq:mar31.14}
\end{equation}

The preceding relations \eqref{eq:Mar31.15}--\eqref{eq:mar31.14} constitute a Lie group
of transformations in the space of the variables $(\rho,u,v,p,e,dx,dy)$
with the infinitesimal generator
\begin{equation}
X=-\rho eq^{2}\Gamma^{2}\partial_{\rho}-pu\partial_{u}-pv\partial_{v}-p^{2}\partial_{p}+(c^{2}p^{2}-e^{2}q^{2})\Gamma^{2}\partial_{e}\label{eq:generator_original2}
\end{equation}
\[
+\left(((c^{2}-u^{2})p+ev^{2})dx-uv(e+p)dy\right)\Gamma^{2}\partial_{dx}+\left(((c^{2}-v^{2})p+eu^{2})dy-uv(e+p)dx\right)\Gamma^{2}\partial_{dy}\ ,
\]
in that they satisfy the Cauchy problem
\begin{equation}
\begin{array}{rcll}
{\displaystyle \frac{d\rho^{*}}{d\epsilon}} & = & -\dfrac{\rho^{*}e^{*}q^{*2}}{c^{2}-q^{*2}}, & \,\,\,\rho_{|\epsilon=0}^{*}=\rho,\\[2ex]
{\displaystyle \frac{du^{*}}{d\epsilon}} & = & -p^{*}u^{*}, & \,\,\,u_{|\epsilon=0}^{*}=u,\\[2ex]
{\displaystyle \frac{dv^{*}}{d\epsilon}} & = & -p^{*}v^{*}, & \,\,\,v_{|\epsilon=0}^{*}=v,\\
{\displaystyle \frac{dp^{*}}{d\epsilon}} & = & -p^{*2}, & \,\,\,p_{|\epsilon=0}^{*}=p,\\[2ex]
{\displaystyle \frac{de^{*}}{d\epsilon}} & = & \dfrac{c^{2}p^{*2}-q^{*2}e^{*2}}{c^{2}-q^{*2}}, & \,\,\,e_{|\epsilon=0}^{*}=e,\\[2ex]
{\displaystyle \frac{d(dx^{*})}{d\epsilon}} & = & {\displaystyle \frac{((c^{2}-u^{*2})p^{*}+e^{*}v^{*2})dx^{*}-u^{*}v^{*}(e^{*}+p^{*})dy^{*}}{c^{2}-q^{*2}},} & \,\,\,(dx^{*})_{|\epsilon=0}=dx,\\
{\displaystyle \frac{d(dx^{*})}{d\epsilon}} & = & {\displaystyle \frac{((c^{2}-v^{*2})p^{*}+e^{*}u^{*2})dy^{*}-u^{*}v^{*}(e^{*}+p^{*})dx^{*}}{c^{2}-q^{*2}},} & \,\,\,(dx^{*})_{|\epsilon=0}=dx.\\[2ex]
\end{array}\label{eq:Mar29.1-1-1}
\end{equation}

Particular invariants associated with the one-parameter class of reciprocal transformations \eqref{eq:Mar31.15} are given by
\begin{equation} \label{c10}
J_{1}=\frac{u}{p},\,\,\,J_{2}=\frac{v}{p},\,\,\,J_{3}=\frac{(cp-eq)(c-q)}{(cp+eq)(c+q)},\,\,\,J_{4}=\rho\frac{J_{3}(c+q)+c-q}{c^{2}-q^{2}}\ .
\end{equation}

\section{Conclusion}

Invariance properties in both nonlinear continuum mechanics and mathematical physics continue to be of research interest. Thus, in particular, invariance under multi-parameter reciprocal transformations in relativistic gasdynamics has recently been established in \cite{art:RogersRuggeri2020,art:RogersRuggeriSchief2020}. Here, a one-parameter subclass of such invariant transformations has been linked to establish Lie group procedures such as set down in the seminal works of \cite{bk:Ovsiannikov1978,ni83}.

Substitution-type invariance principles as originally derived in spatial gasdynamics by Prim \cite{rp49} with extension to magnetogasdynamics (see e.g. \cite{cpcr69}) may likewise be shown to have a Lie group genesis \cite{fo92,foms05}. Such invariance principles are readily established for the relativistic gasdynamics systems of \cite{art:RogersRuggeri2020,art:RogersRuggeriSchief2020}. In magnetogasdynamics, invariance under multi-parameter reciprocal transformations have been derived in \cite{crjkws80} as symmetries of an exterior differential system. In separate developments \cite{cr69,cr72} a range of invariant transformations in non-conducting gasdynamics have been derived via matrix formulations. Conjugation of reciprocal-type transformations has recently been applied in \cite{cr19,crpb20} to solve nonlinear boundary problems incorporating heterogeneity and relevant to seepage soil mechanics. However, a general investigation via Lie group connections of the conjugation and its application of reciprocal-type invariant transformations to such nonlinear boundary value problems of physical importance remains to be undertaken.


\subsection*{Dedication}

This paper is dedicated to the memory of Nail Ibragimov in recognition of his diverse contributions to the application of Lie group methods.

\subsection*{Acknowledgements}

The research of SVM was supported by the Russian Science Foundation Grant No. 18-11-00238 `Hydrodynamics-type equation: symmetries, conservation laws, invariant difference schemes'.

\label{lastpage}

\end{document}